\documentclass[pra,a4paper,aps,twocolumn,superscriptaddress]{revtex4-1}

\usepackage[utf8]{inputenc}
\usepackage[english]{babel}
\usepackage[T1]{fontenc}
\usepackage{amsmath}

\usepackage[pdftex,colorlinks=true,allcolors=blue]{hyperref}

\usepackage{lipsum}

\usepackage {subfigure}
\usepackage{color}
\usepackage {overpic}

\def\bra#1{\mathinner{\langle{#1}|}}
\def\ket#1{\mathinner{|{#1}\rangle}}

\def\vx {\mathbf {x}}
\def\vk {\mathbf {k}}
\def\vq {\mathbf {q}}
\def\vs {\mathbf {s}}
\def\vp {\mathbf {p}}

\def\bra#1{\mathinner{\langle{#1}|}}
\def\ket#1{\mathinner{|{#1}\rangle}}

\def\vx {\mathbf {x}}
\def\vk {\mathbf {k}}
\def\vq {\mathbf {q}}
\def\vs {\mathbf {s}}
\def\vp {\mathbf {p}}
\def\vr {\mathbf {r}} \DeclareMathOperator{\sinc}{sinc}

\begin{document}

\title{Momentum-Resolved and Correlations Spectroscopy Using Quantum Probes}
\date{\today}
\author{Francesco Cosco}
\affiliation{Turku Centre for Quantum Physics, Department of Physics and Astronomy, University of Turku, FI-20014 Turun yliopisto, Finland}
\author{Massimo Borrelli}
\affiliation{Turku Centre for Quantum Physics, Department of Physics and Astronomy, University of Turku, FI-20014 Turun yliopisto, Finland}
\author{Francesco Plastina}
\affiliation{Dipartimento di Fisica, Universit\`a della Calabria, 87036, Arcavata di Rende (CS), Italy}
\affiliation{INFN - Gruppo Collegato di Cosenza, Cosenza, Italy}
\author{Sabrina Maniscalco}
\affiliation{Turku Centre for Quantum Physics, Department of Physics and Astronomy, University of Turku, FI-20014 Turun yliopisto, Finland}
\affiliation{Center for Quantum Engineering, Department of Applied Physics, Aalto University School of Science, P.O. Box 11000, FIN-00076 Aalto, Finland}

\begin{abstract}
We address some key conditions under which many-body lattice
models, intended mainly as simulated condensed matter systems, can
be investigated via immersed, fully controllable quantum objects,
namely quantum probes. First, we present a protocol that, for a
certain class of many-body systems, allows for full momentum
resolved spectroscopy using one single probe. Furthermore, we
demonstrate how one can extract the two-point correlations using
two entangled probes. We apply our theoretical proposal to two
well-known exactly solvable lattice models, a 1D Kitaev chain and
2D superfluid Bose-Hubbard model, and show its accuracy as well as
its robustness against  external noise.
\end{abstract}

\maketitle

\section{Introduction}
\label{sec:sec1} Cold atoms \cite{bloch1,cazalilla1,bloch2} in
optical lattices stand as an almost ideal experimental platform
where several condensed matter models on lattice, such as the
Hubbard and Heisenberg Hamiltonians, can be simulated efficiently
\cite{primadidi,vari,Simulator1}. These simulated systems allow to explore
the properties of  quantum many-body models on lattice in a
protected environment, where lattice imperfections are absent. In
this context it is possible to implement {\it quantum probing
strategies} for many-body systems via a set of controllable and
measurable systems able to extract valuable information. Such
\textit{quantum probes} are conceived as an alternative to more
invasive traditional techniques and their possible use has recently
received a great deal of attention, as experiments
within atomic or spin impurities immersed in optically trapped
atomic gases have  been performed \cite{exp1,expultrafast,singlespin}.

A good body of  literature is already available, showing the
advantages of  such a novel approach when studying optically
trapped cold atoms
\cite{andreas,andreas2,TomiSteve,jaksch2,giammarchi2,strongp,probingBCS}.
Various schemes have been proposed, to probe temperature
\cite{temperature}, phononic excitations \cite {tomihang},
Luttinger physics \cite{recati} and genuine many-body phenomena
such as the orthogonality catastrophe \cite
{plastina1,fischer,schiro,demler}. Quantum  probes
\cite{pinja1,pinja1b,Feynmanprobes} have also been shown to detect critical
phenomena \cite{zwick} and signal phase transitions in trapped
ions \cite{massimo1,massimo2} and spin models
\cite{pinja2,francica}. Other works have employed the quantum
probing paradigm to analyse the spectral properties of complex
quantum networks \cite{network}, reconstruct squeezed thermal
states from an optical parametric oscillator \cite{homodyne} and
estimate the bound to the minimal length of quantum gravity
theories by performing measurements on a harmonic oscillator \cite
{qcom}. The range of interest and applicability of quantum probes
is therefore very wide, although no general theory of quantum
probing has yet been formulated and many questions regarding this
approach remain still unanswered.

Some aspects of such a theory are strictly model-dependent;
nevertheless, one may still wonder whether some general and
model-independent results can be derived. This is the exact aim of
this manuscript. We address two key points of quantum probing from
a more general perspective. First, we discuss a minimal set of
assumptions needed to develop simple and yet general enough
quantum probing protocols; and, second, we analyse what kind of
information regarding a many-body system is accessible via such
protocols. While the first question will naturally lead to
identify some physical systems that are potentially good
candidates for quantum probing, the second focuses more on the
trade-off between the resources needed, such as, e.g.,  the number
of probes needed, and the type of information one can extract. In
what follows, by relying only on some fundamental quantum features
of the probe(s), such as the discreteness of their energy
spectrum, or the initial entanglement between two of them, we are
going to provide efficient tools to detect various properties of a
large class of many-body systems. We aim to investigate two
scenarios. First, we consider a single quantum probe, typically an
impurity of some sort, embedded in a lattice many-body system, and
demonstrate that one can perform momentum resolved spectroscopy of
the many-body system. This is achieved by tailoring the spectrum
of the impurity and measuring transitions probabilities between
its energy levels, with the probe sitting in different positions
with respect to the many-body system. The proposed method does not
give a solution to a previously unsolved task, but its novelty
relies on the different paradigm on which it is based. Contrarily
to standard techniques, indeed, our approach is a good candidate
to perform full momentum resolved spectroscopy in a potentially
less invasive way, avoiding direct measurements performed on the
many-body system. We then move to a two-probe scenario, to show
that the use of entangled probes allows us to monitor the
spreading of correlations throughout the system by measuring one
and two-probe transition rates. So far, entangled probes or
entangled states in general have been mainly employed in the field
of quantum metrology as a resource to improve the accuracy of
parameter estimation \cite {entprobes1,entprobes2,entprobes3,entprobes4}. In this work, instead,
we directly relate the transition rates of entangled probes to
two-point (spatial) correlations characterizing the many-body
system.

As we aim at probing the many-body system in the
least invasive way, we assume weakly coupled probes, so that
transition probabilities between their energy levels can be
expressed in terms of a thermally weighted Fermi Golden Rule.

\section{Quantum Probing Protocols}
\label{sec:sec1}

\subsection{Momentum-resolved spectroscopy}
We start off by setting
the general Hamiltonian of a many-body system interacting with an
impurity probe ($\hbar=1$)
\begin{equation}
\hat H= \hat H_{MB} + \hat H_P + g \hat H_{int},
\label{totalhamiltonian}
\end{equation}
in which $\hat H_{P}=\sum_{\bar n}\epsilon_{n}\ket{\bar n}\bra
{\bar n}$ and  $\hat H_{MB}$ are the free Hamiltonians for the
impurity and for the many-body system, respectively. $\hat
H_{int}$ is the interaction Hamiltonian, which is assumed to
weakly perturb the many-body system while inducing transitions
between the probe energy levels:

\begin{equation}
 \hat H_{int}=\sum_{\epsilon_{\bar m} >   \epsilon_{ \bar n}}  \ket {\bar m} \bra {\bar n} \otimes \hat \Phi {[\bar  m, \bar n]}+ \sum_{\epsilon_{\bar m} <   \epsilon_{ \bar n}}  \ket {\bar m} \bra {\bar n} \otimes \hat \Phi^\dagger{[\bar  m, \bar n]},
 \label{interaction1}
\end{equation}
where $\hat \Phi{[\bar  m, \bar n]} \propto \sum_\vk
\gamma^\vk_{\bar m, \bar n} \hat b_\vk $ ($\hat \Phi^{\dag}$)
describes the single particle absorption (emission) occurring into
the modes of the many-body system with amplitudes $\gamma^\vk_{\bar  m,
\bar n}$, together with the transition  $\ket {\bar n}
\rightarrow \ket {\bar m}$ for the probe. Such an interaction term
is fairly general, and can be related to various models employed
so far to study the non-Markovian dynamics of open systems of non
interacting particles linearly coupled to thermal environments
\cite{nori1}.

A key point, here, is the complete knowledge of the eigensystem of
$ \hat H_P$, which we label $\{\epsilon_{\bar{n}},\psi_{\bar n}
(\vx)\}$, as well as the ability to tune it via some external
control parameters. To move forward in the derivation, we
\textit{i}) select two eigenstates of the probe, namely
$\ket{\bar{g}}$ and $\ket{\bar{e}}$, whose transition frequency
$\nu$ is tunable, \textit{ii)} assume weak coupling between the
impurity and the many body system, \textit{iii}) assume the total
system at the initial time to be of the form $\rho (0) = \ket {\bar g} \bra {\bar g} \otimes
e^{-\beta\hat{H}_{MB}}/\mathcal{Z}_{MB} $, where the many-body system is in
thermal equilibrium, with inverse temperature $\beta$.
\subsubsection {Fermi Golden Rule} Within the above assumptions the total time-dependent transition probability from $\ket
{\bar g}$ to $\ket {\bar e}$ can be written (See Appendix \ref{appendixA} for
the explicit derivation)
\begin{equation}
\begin{split}
\Gamma_{\bar g \rightarrow \bar e}(t)=  g^2  \int_0^t \, \mathrm dt_1    \int_0^t \, \mathrm dt_2 \langle  \hat \Phi^\dagger {[\bar  g, \bar e]}(t_1)  \hat \Phi {[\bar  e, \bar g]}(t_2)\rangle \\
   e^{-i \nu (t_1-t_2)}+O(g^{4}).
\end{split}
\label{rate1}
\end{equation}
For the sake of concreteness we make two further assumptions usually satisfied in experiments with cold
atoms.
We focus our attention on \textit{iv}) systems
characterized by a (known) lattice structure.
Formally, this allows us to characterize our system in terms of Bloch
functions $w_{\vk}(\vx)$, with corresponding frequency
$\omega_\vk$ and ladder operators $\hat{b}_{\vk}$. We also make
the standard assumption of confining the dynamics to the lowest
Bloch band. Any Bloch function, in turn, can  be expanded in terms
of site-localised Wannier functions as $w_\vk (\vx)= \sum_\vr
\gamma_\vk e^{i \vk \vr} W_\vr (\vx)$. 
We also the assume the amplitudes $\gamma^\vk_{\bar  m, \bar n}
$ to be proportional to the overlapping integrals,
$\gamma^\vk_{\bar  m, \bar n}  = \int \mathrm d \mathbf x \;
\psi_{\bar e}^* (\vx) \psi_{\bar g} (\vx) w_{\vk} (\vx)$, where
$\psi_{\bar {e}/\bar{g}}(\vx)$ are the probe wavefunctions.

Finally, \textit{v}) we assume the probe to be local,
so that the interaction is localised on one lattice site, say
${\bf 0}=(0,0,0)$, and the only relevant overlapping integral
is that one involving the corresponding Wannier function,
$W_{{\bf 0}}(\vx)$.

The time-rescaled transition rate then reads
\begin{equation}
\begin{split}
\tilde{\Gamma}_{\bar g \rightarrow
\bar e}(t) &\equiv\Gamma_{\bar g \rightarrow \bar e}(t)/g^{2}t^{2}  \\ &=\sum_\vk |J_\mathbf{0}  \gamma_\vk |^2
  \sinc^2\left[\frac {(\nu -\omega_\vk)t}{2}\right] n_{\vk},
\label{rate2}
\end{split}
\end{equation}
in which $J_\mathbf{0} = \int \mathrm d \mathbf x \;  \psi_{\bar
e}^* (\vx) \psi_{\bar g} (\vx) W_{0} (\vx)$ and $n_{\vk}$ is the
average number of thermal excitations at $\omega_\vk$. When
measuring the transition rate Eq.~\eqref{rate1} as a function of
the probe frequency, one will observe resonance peaks revealing
the (single-particle) excitation spectrum of the many-body system.
The amplitudes of such peaks read,  according to
Eq.~\eqref{rate2}, $A_\vk^2 =d_\vk |J_\mathbf{0} \gamma_k |^2
n_{\vk}$, where $d_\vk$ is the $\vk$-th mode degeneracy.

The position of the peaks gives a direct measure of the excitation
frequencies of the system; but, in order to fully reconstruct the
dispersion relation of the many-body system, the correct $\vk$ has
to be associated to each $\omega_{\vk}$. We have found that this
can be achieved by repeating the protocol while changing the
position of the probe and comparing the corresponding transition
rates. In particular, we show below through some specific examples
that the number of different measurements that we need to perform
is related to the dimensionality of the system to be probed;
therefore we label the
different measurement sets with the index $i=1,\ldots d$. 
The ratio between the amplitudes of the resonant peaks obtained from these extra
measurements and those of the first one is found to satisfy the relation
\begin{equation}
 \frac{{A^2_\vk}^{i}}{{A^2_\vk}}=\left|\frac{J_{\mathbf i}}{J_{\mathbf{0}}}\right|^{2} \sum_{\mathbf k | \nu=\omega_{\mathbf k}} G_i(\vk), \quad i=1,...,d
 \label{rateasintotico2}
 \end{equation}
where $J_{\mathbf {i}}= \int \mathrm d \mathbf x \; \psi_{\bar
e}^* (\vx-{\mathbf a_{i}}) \psi_{\bar g} (\vx-{\mathbf a_{i}})
W_{0} (\vx)$ and $G_i(\vk)$ are functions of the momentum that can
be deduced from geometric considerations (see Appendix \ref {appendixA}). In order to observe the resonant peaks the time of measurement should be larger then the
typical time scale associated to the low-energy portion of the
spectrum and, on the other hand, short enough to guarantee the
validity of the perturbative approach. An optimal choice of this time lies in the range
$\bigg [\textrm{max}\left[\frac{4\pi}{\vec{v}_{\vk}\cdot\vec{V}_{R}}\right],1/g \bigg ]$,
in which $\vec v_\vk$ and $\vec{V}_{R}$ are the group velocity and
the volume of the first Brillouin zone. For instance, in the
simple case of phonons in a 1-D lattice, the lower bound is $4 \pi
N_s a/c_{s}$, with $c_{s}$ being the speed of sound.

Once the ratio $J_i/J_0$  is measured,
Eqs.~\eqref{rateasintotico2} should be inverted to associate each
momentum $\vk$ to the corresponding frequency $\omega_{\vk}$. This
procedure will be exemplified
in the following section, where we apply the protocol
to two different physical models, and show its properties and study its
robustness against noise.

\subsubsection {Master equation and thermodynamic limit} The protocol, as outlined above, appears not to be practically feasible if the the many-body
system under scrutiny is too large. Indeed, in this case, the
resonance peaks would become increasingly closer, making the probe
tunability rather challenging if not impossible. However, in the
thermodynamic limit and for a gapless energy spectrum,
Eq.~\eqref{rate2} still does give the correct transition rates
provided that the sum is replaced by an integral, in the continuum
limit $\sum_\vk \omega_\vk \rightarrow \int d \vk $. In this case,
using standard open quantum system approaches within the
Born-Markov approximation, one can derive a master equation
describing the dynamics of the probe. Assuming the probe is
resonant with some excitation energy of the system, $\nu=\omega
(\vk)$, we have

\begin{widetext}
\begin{equation}
\begin{split}
\frac {d\rho_P} {dt} =g^2 |J_0 \gamma (\vk)|^2\Bigg [\left[1+sn(\nu) \right] \left(\sigma^- \rho_P \sigma^+-\frac {1}{2} \left \{ \sigma^+\sigma^-, \rho_P \right \}\right)
+   n(\nu)  \left(\sigma^+ \rho_P \sigma^--\frac {1}{2} \left \{ \sigma^-\sigma^+, \rho_P\right \}\right)\Bigg],
\label{mastereq}
\end{split}
\end{equation}
\end{widetext}
in which we have neglected the Lamb and Stark contributions,
$s=\pm 1$ distinguishes between bosonic/fermionic baths and  $\hat \sigma^{\pm}$ are the two-level system ladder
operators.  
When the
probe is initialised to $\rho_P (0)= \ket {\bar g} \bra { \bar
g}$, simple solutions are found for both  bosonic and fermionic
environments respectively
\begin{align}
\rho_{\bar e \bar e} (t) &= \frac {n(\nu)}{2 n(\nu)+1} (1-e^{-\gamma_B t}), \\
\rho_{\bar e\bar e} (t) &=  {n(\nu)} (1-e^{-\gamma_F t}),
\label {stationary}
\end{align}
in which $\gamma_B =g^2 d(\vk) |J_0 \gamma(\vk)|^2 (2 n(\nu)+1)$
and  $\gamma_F =g^2 d (\vk) |J_0 \gamma (\vk)|^2$. The information
about the momentum is encoded into these decay rates,
and the ratio needed in  Eq.~\eqref{rateasintotico2} can be
obtained by means of an extra set of measurements, as outlined in the previous paragraph.
\subsubsection {Bloch functions spectroscopy}
We conclude this section by showing that the scheme discussed so
far also allows, at least in principle, for the reconstruction of
the Bloch functions, provided that the probe can be placed at  a
varying distance $\vs$ from its initial position. Indeed, for a
displacement $\vs$ of the probe, the amplitude $|A_\vk|$ of the
resonant peak reads
\begin{equation}
\frac {|A_\vk|}{(d_\vk  n_\vk)^{\frac{1}{2}}} =\Bigg|\int \mathrm d \mathbf x \;  \psi_{\bar e}^* (\vx-\vs)  \psi_{\bar g} (\vx-\vs)  w_{\vk} (\vx)   \Bigg|= |\psi   * w_\vk | (\vs)
\label {blochsp}
\end{equation}
where $\psi (\vx-\vs)\equiv   \psi_{\bar e}^* (\vx-\vs) \psi_{\bar
g} (\vx-\vs)$, and $*$ denotes the convolution integral. If both
$\psi$ and $w_\vk$ are real functions, then the Fourier transform
of  Eq.~\eqref{blochsp} gives $A_\vk (\vp) = \psi (\vp)
w_\vk(\vp)$. Since the eigenstates of the probe are known, it is
possible to extract  $ w_\vk( \vp)$ from the measurements. By
transforming back to real space, the Bloch function $ w_\vk (\vx)$
can be finally obtained. Notice that, due to the symmetry of the
lattice, $A_\vk (\vs)$ needs to be sampled in a fraction of the
$1^{\textrm{st}}$ Brillouin zone only.

\subsection{Probing of Quantum Correlations}
In the previous section we showed that a single quantum impurity
with a discrete and tunable energy spectrum allows for a complete
reconstruction of the dispersion relation of a certain class of
many-body systems. Having more than a single controllable probe at
our disposal, and assuming that entangled states can be prepared,
correlation properties of the many-body system can be extracted.

In what follows, we show how the spreading of two-point
correlations in the many-body system can be mapped onto a simple
function of the impurity transition rates in a two-entangled-probe
setting. Let us assume that two identical impurities, say $A$ and
$B$, are placed on $\vx_{A}$ and $\vx_{B}$ respectively. The total
system+probe Hamiltonian reads
\begin{equation}
\hat H=\hat H_E  + \hat H_{P_A}+\hat H_{P_B}+g \hat H_{int},
\label{totaham2imp}
\end{equation}
with an interaction Hamiltonian analogous to the one used above,
\begin{equation}
\begin{split}
\hat H_{int}= \sum_{\bar n, \bar m}  \ket {\bar n}_A \bra{\bar m}_A \otimes \hat \Phi_{\vx_A}^{[\bar m_A,\bar n_A]} \\ +\sum_{\bar n,\bar m}  \ket {\bar n}_B \bra{\bar m}_B \otimes \hat \Phi_{\vx_B}^{[\bar m_B,\bar n_B]},
\label{interaction2imp}
\end{split}
\end{equation}
in which $\hat \Phi_{\vx_{A|B}}^{[\bar  m, \bar n]}$ are now the
many-body observables whose correlation function we are interested in. We
consider two-level impurities prepared in the Bell state $\ket
{\Psi_{0}}=\frac {1}{\sqrt 2}(\ket {\bar g_A ,\bar e_B }+\ket
{\bar e_A, \bar g_B })$. The combined initial state of the
impurities and of the many-body system is therefore $\rho (0) =
\ket {\Psi_{0}}\bra {\Psi_{0}} \otimes
e^{-\beta\hat{H}_{MB}}/\mathcal{Z}_{MB}$. As we demonstrate in detail in Appendix \ref{appendixB}, the spreading of
correlations in the many-body system following the embedding of
the impurities are captured by the following combination of one-
and two-impurity transition rates, that we compute as in
Eq.~\eqref{rate1}:
\begin{equation}
\begin{split}
\overline \Gamma &= \Gamma^{(2)}_{\ket{\Psi_{0}} \rightarrow \ket {\bar e_A ,\bar e_B} }-\sum_{P=A,B} \frac{1}{2}\Gamma^{(1)} _{\ket {\bar g_P} \rightarrow \ket {\bar e_P} } \\ 
&=\frac {g^2}{2} \int_0^t dt_1  \int_0^t dt_2  \sum_{\substack{i,l=A,B\\i \neq l}} \langle  \hat \Phi_{\vx_i}^{[\bar e_i,\bar g_i]}(t_1)\hat \Phi_{\vx_l}^{[\bar g_l,\bar e_l]}(t_2) \rangle \\ &e^{i\nu(t_1-t_2)}.
\label{rate2probes}
\end{split}
\end{equation}
By tracking the time-evolution of $\overline \Gamma_{}$, two-point
correlations can be monitored.  Given its model-independence, this
protocol can be also applied  to many-body systems with
long-range interactions, which break the Lieb-Robinson bound \cite {LBbound1,LBbound2}.

\section{Applications}
In this section we apply the one- and two-probe
protocols described above to two examples. First we consider
probing a 1D long-range fermionic hopping model that has recently
attracted an increasing interest as its exact solvability makes it a good candidate to explain
qualitatively the physical behaviour behind the propagation of correlations in systems with long range
interactions \cite{kitaev1}; then we discuss the probing of a
2D Bose Hubbard model in the superfluid phase. In particular, in
the second example, the experimental implementation of the
momentum-resolved spectroscopy protocol in a cold atom platform is
discussed in some details, and compared to other available
methods.

\subsection{ 1D Kitaev chain} Recently, long-range hopping models have
been receiving a renewed attention due to their experimental
realisability, demonstrated in solid state systems. As an example
the so-called helical Shiba chains made of magnetic impurities on
an s-wave superconductor have been implemented
\cite{expki1,expki2}. On the other hand these models share
some features with long range Ising models, experimentally realizables
with trapped ions and cold atoms \cite {isisimu1,isisimu2}, where also long
range tunnelling has been observed \cite {lrtunnelling}.

\begin{figure*}[!t]
 \centering
 \textbf {(a)} \qquad\qquad\qquad\qquad\qquad\qquad\qquad\qquad\qquad \textbf {(b)}\par \medskip

   { \hspace{-10 mm} \includegraphics[scale=0.5]{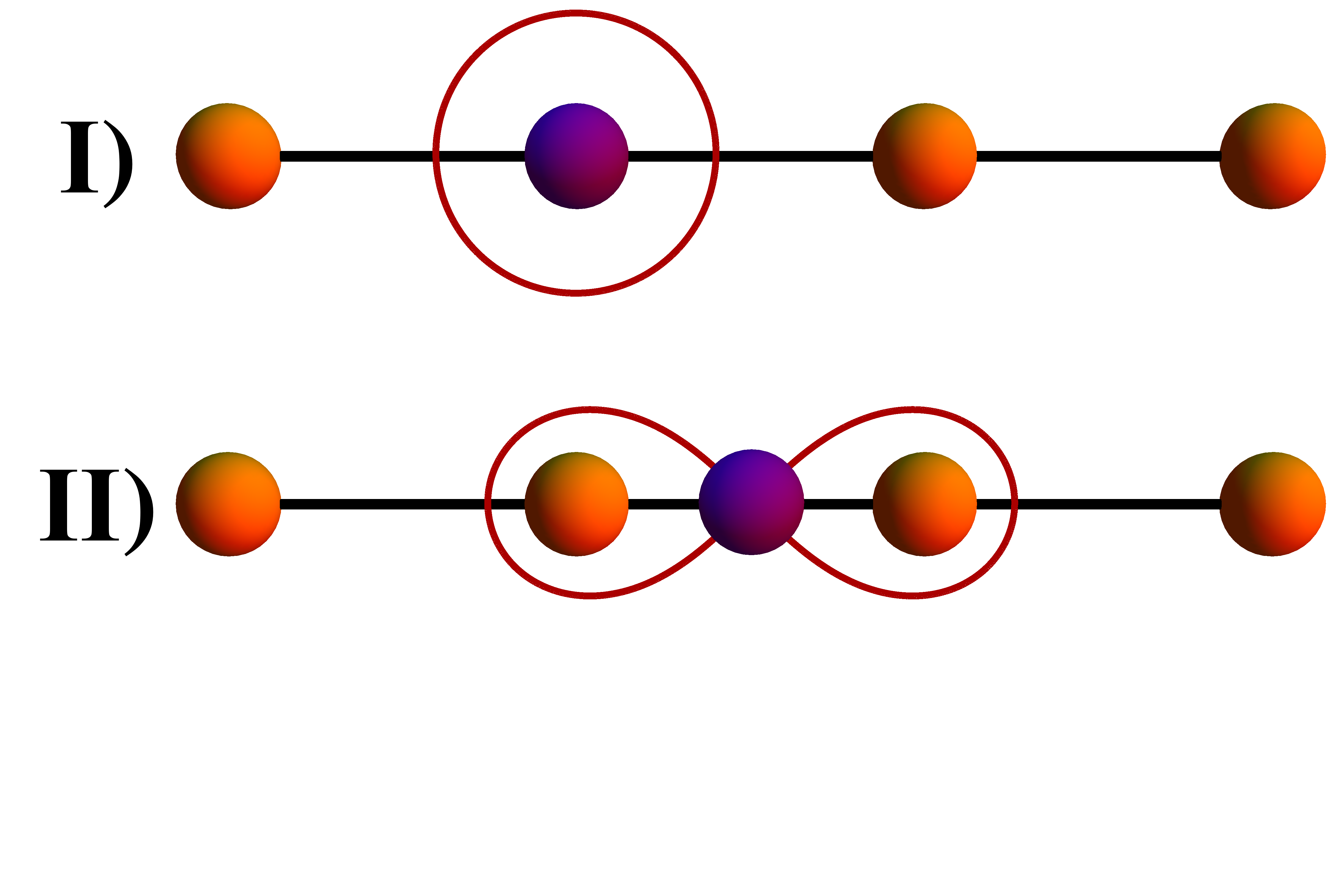}}
 \hspace{10 mm}
   {\includegraphics[scale=1]{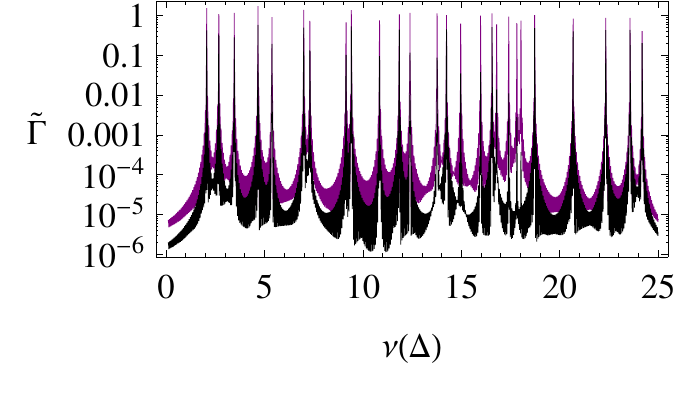}}

 \textbf {(c)} \qquad\qquad\qquad\qquad\qquad\qquad\qquad\qquad\qquad \textbf {(d)}\par 
\vspace{-4 mm}
 \subfigure
   {\includegraphics{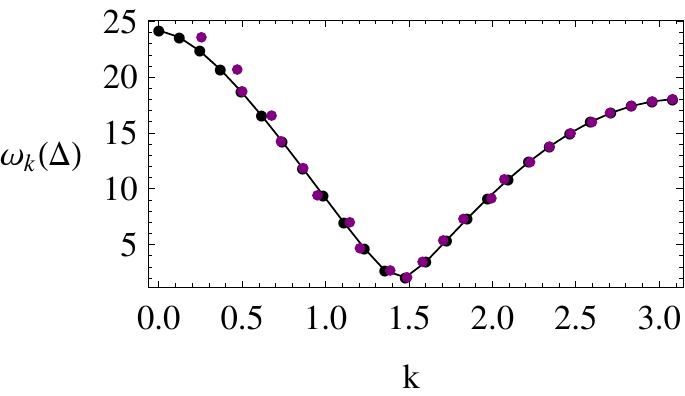}}
 \hspace{14 mm}
\subfigure
   {\includegraphics[scale=0.95]{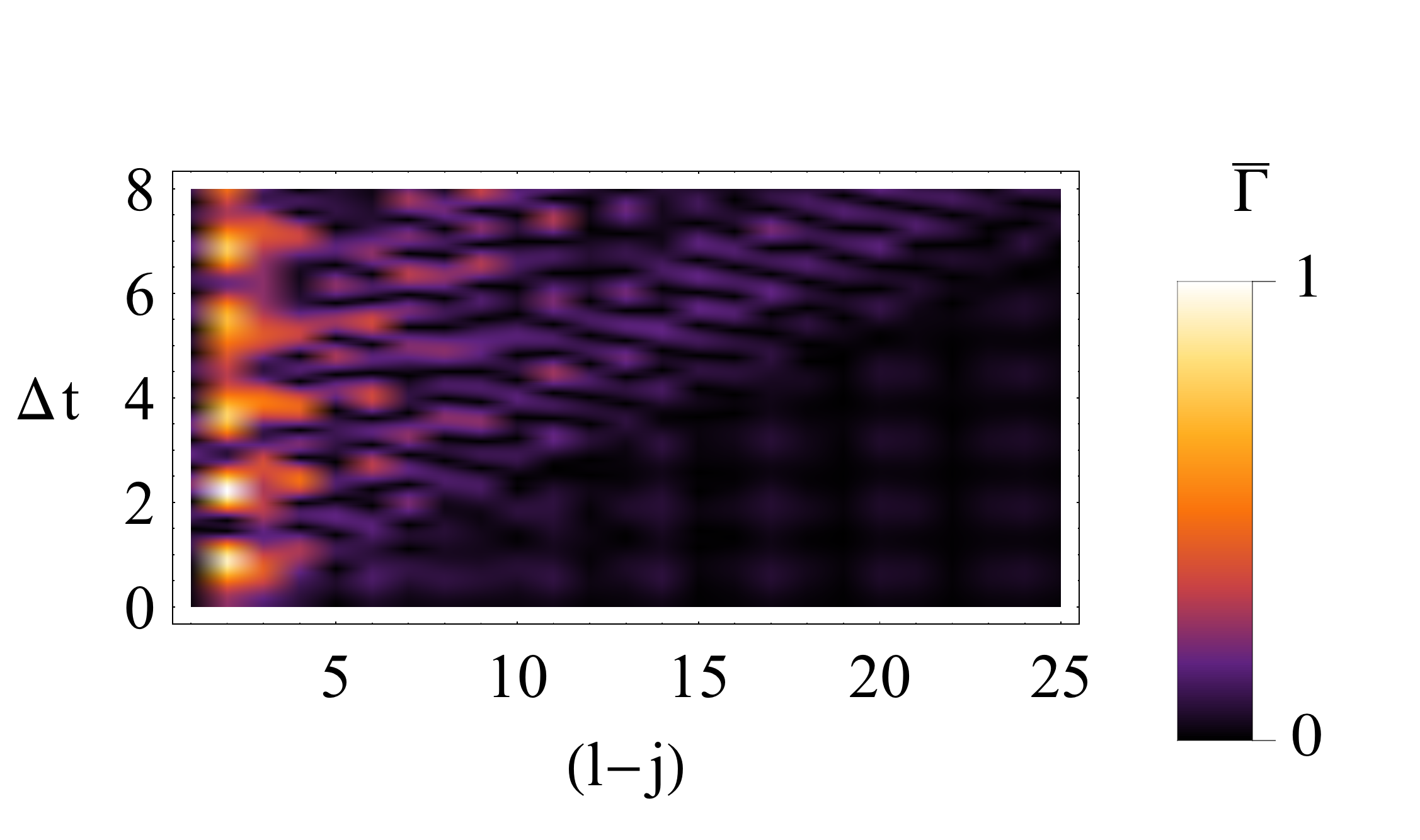}}
\caption{(a) Sketch of a probe overlapping with
either one or two lattice sites of a 1-D chain. Measurements in
both of the positions are required to perform momentum resolved
spectroscopy. (b) Transition probabilities corresponding to the
two probe positions, displayed in black and purple, respectively. (c)
Reconstructed spectrum. The purple dots are the frequencies extracted
from the transition probabilities data, while the black ones are
the exact values. Here the source of error is given by the finite
sampling in the probe frequency $\nu$. An additional random $2\%$
error has been added to the ratios of the peaks, and $g'/g$ has been estimated by summing over  them. The
agreement is very good, although two points are missing in the
dispersion relation. The corresponding frequency is clearly
visible in the transition probability, but the errors hinders the
estimation of the proper momentum. (d) Quantum correlation
function with a clear light-cone structure emerging  while it
spreads across the lattice. All  the plots are drawn for
$J/\Delta = 5$, $\alpha=0.3$, and $N_s=51$. Everywhere, $\Delta$ is
the energy and inverse time unit.} \label{kipic}
\end{figure*}

Here, we consider a 1D
lattice model for spinless fermionic excitations to prove both the usefulness
and the robustness of our protocols. The many-body Hamiltonian has
the form of a generalized Kitaev ring with long-range hopping
\cite {kitaev0, kitaev1,kitaev2}
\begin{equation}
\hat H = \sum_{l\neq j=1 }^{N_s} J_{lj} \hat c^\dagger_l \hat
c_j+\Delta \sum_{j=1 }  \hat c^\dagger_j \hat c_{j+1}^\dagger +
\mathrm {H.c.},
\end{equation}
in which $\Delta$ and $J_{lj}=J |\pi/(N_s \sin
[\pi(l-j)/N_s])|^{\alpha}$ are the pairing and long range
tunnelling coefficients, respectively. We choose a two-level
probe, $\hat H_P= \nu \hat S_z $, for which
we assume the following interactions for the first and second step of the momentum resolved spectroscopy,
\begin{align}
\hat H^I_{int} &=g \hat \sigma^+ \sum_k \cos (\theta_k/2) \hat {\tilde c}_k+ \textrm{H.c.},
\label{kitaevint}\\
\hat H^{II}_{int} &=g \hat \sigma^+ \sum_k [1+ \cos (k)]\cos (\theta_k/2) \hat {\tilde c}_k+  \textrm{H.c.}.
\label{kitaevintb}
\end{align}
These Hamiltonians are precisely of the form given in the general
formulation above, with $ \hat \Phi{[\bar e, \bar g]} = \hat \Phi
{[1, 0]} = \sum_k \cos (\theta_k/2) \hat {\tilde c}_k$.
Furthermore, $\hat H^I_{int}$ and $\hat H^{II}_{int}$ correspond
to the two measurement configurations required to achieve momentum
resolved spectroscopy in 1D, as illustrated in Fig.~\ref{kipic}.
The energy and momentum resolved spectrum extracted from the
transition rates are displayed in Fig. \ref {kipic}, where we also
assumed a non-perfect control of the coupling constants, thus
introducing some systematic error (see caption). These plots
demonstrate the accuracy and robustness of the protocol for the
momentum resolved spectroscopy, showing that it is possible to
reconstruct a non-monotonic (but non-degenerate) energy spectrum.

As for the probing of correlations, we consider two probes placed
on sites $l$ and $j$, and assume now an explicit
position-dependent interaction. We make the replacement $\sum_k
\cos (\theta_k/2) \hat {\tilde c}_k \rightarrow \hat c_l$, in
order to get a clearly position-dependent observable. Nonetheless,
the previous interaction can be obtained from this new one if a
rotating wave approximation is applicable.

The $\overline{\Gamma}$ function then reads
\begin{equation}
\overline \Gamma_{} =\frac {g^2}{2} \int_0^t dt_1  \int_0^t dt_2   \langle \hat c_l^\dagger(t_1)\hat c_j(t_2)+\hat c_j^\dagger (t_1)\hat c_l(t_2) \rangle e^{i \nu(t_1-t_2)}.
\end{equation}
The time-evolved correlations, as extracted from $\overline
\Gamma$, are also displayed in Fig. \ref {kipic}.

\subsection{ Bose-Hubbard: Superfluid 2D }

\subsubsection{Bosonic environment and mean field theory}
Before moving to the second application of our protocol,  we first
show, in general terms, how the momentum resolved protocol can be
applied in the experimentally relevant case of an interacting gas
of cold Bosons. The starting point is to consider that also the
probe will interact with the Bose gas via a density-density
interaction, which can be described with the usual assumption of a
contact potential. Denoting by $\hat \Phi(\mathbf x)$ the
many-body field operator, we can write
\begin{equation}
 \hat H_{int}= \sum_{\bar n, \bar m} \int \mathrm d \mathbf x \;  \psi_{\bar m}^* (\vx)  \psi_{\bar n} (\vx)  \ket {\bar m} \bra {\bar n} \otimes \hat  \Phi^\dagger (\vx) \hat \Phi (\vx).
\end{equation}
Loosely speaking, with such a probe-system interaction, the probe
transition probability will show a resonance every time the energy
difference between its levels matches the energy difference
between the modes of the many body system (provided the transition
is allowed, thanks to a non-vanishing overlapping integral).

By adopting a mean-field description for the lattice many-body
system, the field operator can be written as $\hat \Phi (\vx)
=\Phi_0 (\vx) + \hat {\delta \Phi} (\vx)$, with $\hat {\delta
\Phi} (\vx)$ describing linear fluctuations around the mean-field
value $\Phi_0 (\vx)$. Given the lattice structure of the many-body
system, we can expand $\delta \Phi
(\vx)=\sum_{\vk}w_{\vk}(\vx)\hat{b}_{\vk}$, in which
$w_{\vk}(\vx)$ are lattice Bloch functions. Neglecting
contributions that are quadratic in the fluctuations, the
interaction Hamiltonian reads
\begin{equation}
\begin{split}
 \hat H_{int} &= \sum_{\bar n, \bar m} \int \mathrm d \mathbf x \;  \psi_{\bar m}^* (\vx)  \psi_{\bar n} (\vx)  \ket {\bar m} \bra {\bar n} \otimes \\ & \Bigl\{ \Phi_0^2  +  \Phi_0 \sum_{\vk}\left[ w_\vk (\vx)^* \hat b_\vk^\dagger +w_\vk (\vx) \hat b_\vk\right]\Bigr \}.
\end{split}
\end{equation}
Using the same preparation and measurement procedures discussed in
general terms in Sec. \ref{sec:sec1}, the rescaled transition rate
analogous to that in Eq. (\ref{rate2}) reads
\begin{equation}
\Gamma_{\bar g \rightarrow \bar e}(\nu,t)=g^2\Bigl\{\Gamma_0(\nu,t)
+\sum_k \Gamma^-_\vk(\nu,t)+ \Gamma^+_\vk(\nu,t) \Bigr \}\, ,
\end{equation}
where
\begin{align}
\Gamma_0(\nu,t) &=|\gamma_0|^2 \Phi_{0}^{4} \sinc(\nu t),  \\
\Gamma^-_\vk(\nu,t)&=\Phi_{0}^{2} |J_0  \gamma_\vk |^2
  \sinc^2\left[\frac {(\nu +\omega_\vk)t}{2}\right] (1 + n(\omega_\vk)), \\
\Gamma^+_\vk(\nu,t)&= \Phi_{0}^{2} |J_0  \gamma_\vk |^2
  \sinc^2\left[\frac {(\nu -\omega_\vk)t}{2}\right] n(\omega_\vk).
\end{align}
with $\gamma_0= \int \mathrm d \mathbf x \;  \psi_{\bar e}^* (\vx)
\psi_{\bar g} (\vx)$. For large enough measuring times, resonant
peaks appear in the $\nu$-dependent transition rates, whose
amplitudes read $A_\vk^2 =d_\vk |J_\mathbf{0} \gamma_k
|^2n_{\vk}$, if $\nu>0$.

By repeating the measurement with the probe in different positions,
as described in Sec. \ref{sec:sec1}, the analogous of
Eq.~\eqref{rateasintotico2} can be obtained, from which the
momentum resolved spectrum can be finally extracted, as discussed
in detail below for the case of a 2D Bose-Hubbard model.
\subsubsection{2D Superfluid}
Let us now present the second application of our protocol. To this
end, we consider a gas of cold bosonic atoms in the lowest energy
band of a 2D optical potential. This system is very well described
by the Bose-Hubbard model
\cite{hubbard,hubbard2,kanamori,jaksch1}:
\begin{equation}
\hat H =-J \sum_{\langle l, j \rangle}  \hat c^\dagger_l \hat c_j + \frac {U} {2} \sum_j \hat n_j (\hat n_j-1)-\mu \sum_j \hat n_j.
\label{BHH}
\end{equation}
In the limit $J\gg U$, the system is in the superfluid phase, with
a low-energy spectrum due to phononic excitations above a uniform
Bose-Einstein condensate \cite {oosten1}. These excitations have
been successfully resolved in energy using techniques such as
magnetic gradients \cite{bloch3} and lattice depth modulation
\cite{OPL}. Furthermore, a full momentum-resolved spectroscopy has
been performed using two-photon Bragg spectroscopy in \cite
{BRAGG,BRAGG2}. Although quite successful, such methods strongly
interfere with the dynamics of the gas and are therefore very
invasive. We apply our quantum probe protocols in this case, by
assuming the probe immersed in the lattice to be an atomic quantum
dot trapped in a 3D harmonic potential, with energy states
$\psi_{\bar n} (\mathbf x)= \psi_{n_x}^{(\nu_x)}
(x)\psi_{n_y}^{(\nu_y)} (y)\psi_{n_z}^{(\nu)} (z)$. As shown in
Ref. \cite {jaksch2}, the density-density interaction typical of
cold atoms, when applied to the case of a superfluid/BEC in the
mean-field approach, gives rise to a linear coupling of the
impurity to density fluctuations in the many-body system. 

To start the reconstruction procedure, we evaluate the transition
probability between impurity states along the direction orthogonal
to the lattice, say, e.g., the $z$ axis,  $\tilde \Gamma_{\bar 0
\rightarrow (0,0,n_z) }$ as a function of the probe trapping
frequency in that direction, $\nu_z$. As a matter of fact, $\tilde
\Gamma_{\bar 0 \rightarrow (0,0,n_z) }$ depends on $\nu_z$ as well
as on the overlap between the lattice Wannier states and the
unperturbed eigenfunctions of the impurity. For the first
measurement, we find that the transition rate is given by exactly
the same expression reported in Eq. (\ref{rate2}), but with the
prefactor now given by
\begin {widetext}
\begin{equation}
\begin{split}
J_0 \gamma_k =\int dx \, \psi_{n_x=0}^2(x)W_0^2(x) \int dy \, \psi_{n_y=0}^2(y)W_0^2(y) (-1)^{n_z}\sqrt {
{m}}\frac {\gamma_{n_z}^{1/2}\gamma_0^{1/2}}{\pi}\sqrt {\nu_z} \beta_\vk,
\label{gammaA}
\end{split}
\end{equation}
\end{widetext}
\begin{figure*}[!t]
\centering
\textbf {(a)}\par
\includegraphics[scale=0.90]{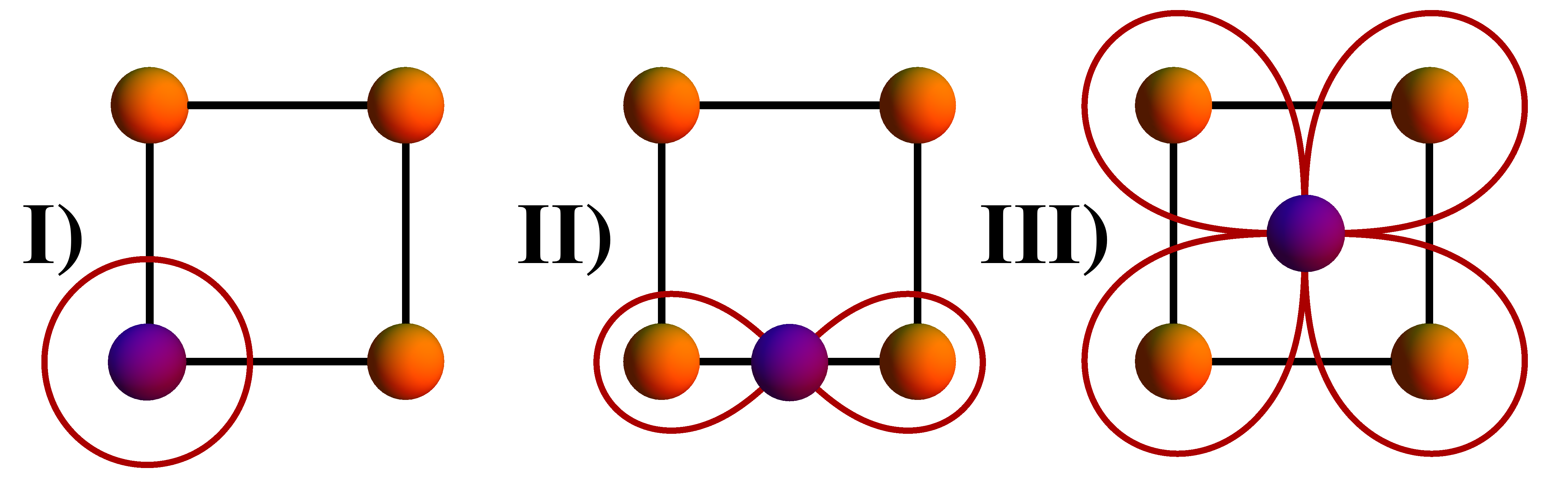}

 \textbf {(b)} \qquad\qquad\qquad\qquad\qquad\qquad\qquad\qquad \textbf {(c)}\par \medskip
\includegraphics[scale=1]{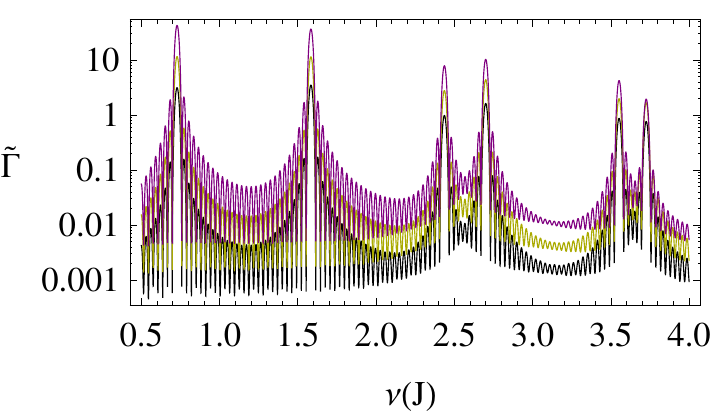}
\includegraphics[scale=1]{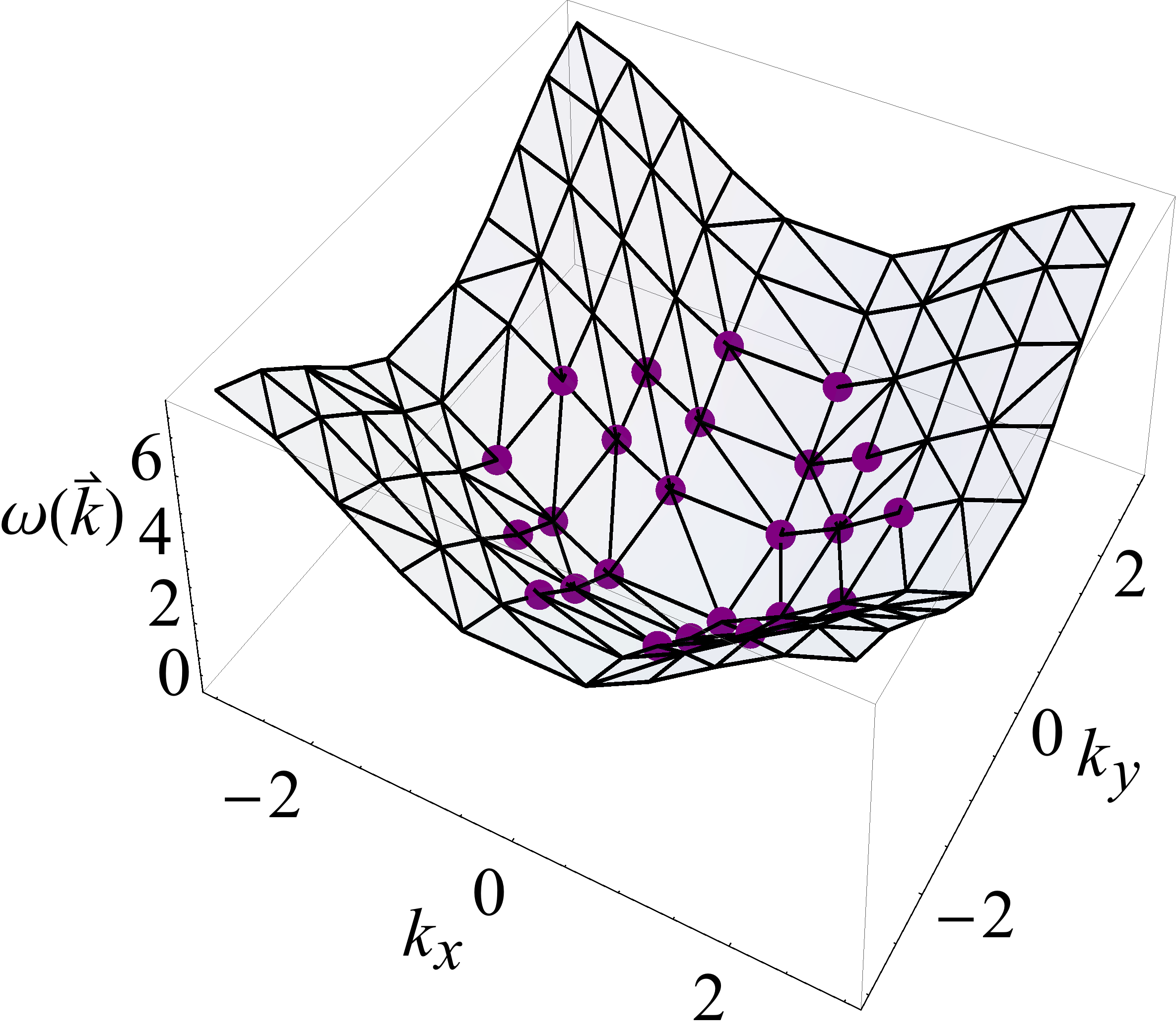}

\textbf {(d)}\par \medskip

\includegraphics[scale=1]{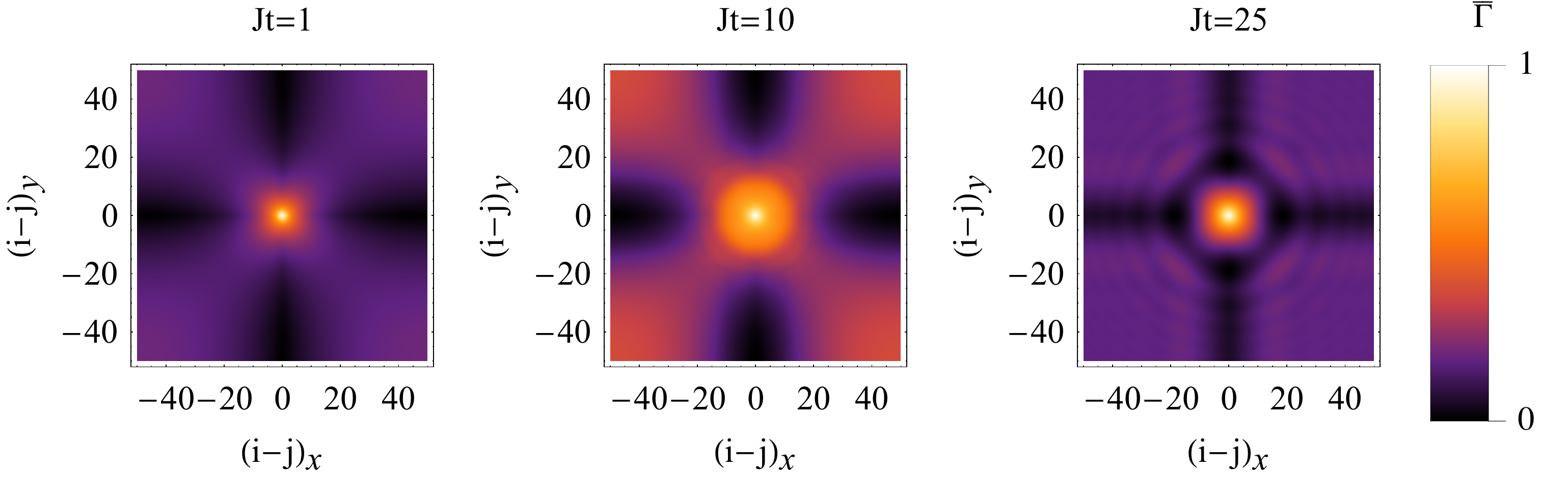}

\caption{(a) Sketch of the probe-positions in the
three stages of the protocol. (b) Semi-log plot of the  low-energy
transition probabilities $\tilde \Gamma^{(\mathrm  I, \mathrm
{II}, \mathrm {III})}$, rescaled by $|\frac {J_i}{J_0}|^2$, for
the three cases shown in (a), drawn in black, yellow and purple,
respectively for a lattice of $N_s=121$ sites. (c) Reconstructed
dispersion relation (purple dots). (d) The correlation function
defined in Eq. \ref {density} for a square lattice of $N_s=121^2$
sites, displayed at three different times. In each plot, the value
of the correlation function has been scaled to its maximum, $J$
has been taken as energy and inverse time units, with $U=0.1 J$.}
\label {sflpic}
\end{figure*}

As depicted in Fig. \ref{sflpic}, and as discussed briefly in
general terms above, the reconstruction of the single particle
excitation spectrum for a 2D lattice system requires two further
measurements, with the probe placed in different positions. Thus,
the overall protocol develops in three subsequent steps with the
probe placed as in in Fig. \ref{sflpic} (a). {\bf I)} ,
{\bf II)} and {\bf III)}, during which the rates
$\Gamma^{\mathrm {I}}, \Gamma^{\mathrm {II}}$ and $\Gamma^{\mathrm
{III}}$ are extracted.

In a realistic experimental realization the in-plane trapping
frequencies, $\nu_x$ and $\nu_y$, can be modified in order to
achieve a suitable wave-function overlap with the nearest
neighboring sites, as necessary for the protocol. The information
obtained by using these three steps allows for the reconstruction
of the full dispersion relation $\omega(\vk)$. Indeed, by taking
the ratios of the second and third transition rates with respect
to the first one at each resonant peak, one obtains two relations
that can be inverted to extract the two components of the momentum
vector $\vk = (k_x,k_y)$ associated to the selected peak
frequency. In our case, we have (see Appendix \ref{appendixA} for
the details)
\begin {equation}
\begin{cases}\frac {\tilde \Gamma^{\mathrm {II}}}{\tilde \Gamma^{\mathrm {I}}} = 2 |\frac {J_1}{J_0}|^2 \left [\cos^2 (\frac {k_x a} {2} )+\cos^2 (\frac {k_y a} {2}
)\right ],
\\ \frac {\tilde \Gamma^{\mathrm {III}}}{\tilde \Gamma^{\mathrm {I}}} = 16  |\frac {J_2}{J_0}|^2 \, \cos^2 (\frac {k_x a} {2} )\cos^2 (\frac {k_y a} {2} ). \end{cases}
\end{equation}
These relations are easily numerically inverted, so that a vector
$\vk$ is associated to each peak frequency $\omega$ found in the transition rates, with the result reported in Fig.
\ref{sflpic}.

Finally, for measuring the correlation function, one needs to use two entangled probes, located at sites $l$ and
$j$. Following the line of the derivation given above, one obtains the
collective decay rate $\overline{\Gamma}$ analogous to that of Eq.
(\ref{rate2probes}). For this system we show in detail in Appendix \ref {appendixB} that the following expression holds:

\begin{equation}
\begin{split}
\overline \Gamma_{} &=\frac {\bar g^2}{2} \int_0^t dt_1  \int_0^t dt_2   \langle \hat n_l(t_1)\hat n_j(t_2) + \\ &+\hat n_j (t_1)\hat n_l(t_2) \rangle e^{i \nu(t_1-t_2)},
\label{density}
\end{split}
\end{equation}
where $\bar g$ is a new coupling constant containing the overlap
integral. Due to the density-density interaction of the impurity
with the boson gas, thus, the function $\overline{\Gamma}$
contains information about the density-density correlation
function, whose behaviour can be extracted from the measured
values of the transition rates, as reported in Fig. \ref{sflpic}.

 {Both the one- and two-probe protocols are in principle feasible for a 2D superfluid with current technology \cite {exp1,exp2,ott,osp}. In particular, it has been proven that the control of the position of an impurity in an optical lattice can be achieved in different ways. For example using species-selective or spin-selective optical lattice, by changing the polarization angle of the laser beams generating the lattice it is possible to obtain two spin dependent potentials that split in opposite direction \cite {cohetransport}. Efficient control over a single spin at a specific site of an optical lattice has been achieved recently, with the further benefit  of leaving the atom in the vibrational ground state, that is the initial state of our probe \cite{singlespin}. Detecting the locations of such impurities is nowadays in the grasp of the experimentalists \cite {micro1,micro2} as well as measuring the populations of excited vibrational states  \cite {motmeas1,motmeas2}.}
\section{Conclusions } We have demonstrated how global
properties of a certain class of many-body systems can be
extracted by means of controllable quantum probes. Our approach is
rather general, being only based on the assumptions of  full
control of the probe position and of its interaction with the
many-body system. Within such framework, we have demonstrated that
it is possible to obtain the single particle excitation spectrum
of the system and perform momentum-resolved spectroscopy via
energy-resolved measurements on a single quantum probe. Moreover,
the spatial profile of the unperturbed Bloch functions can be
reconstructed. By exploiting entangled probes that locally alter
the equilibrium configuration of the many-body system, it is
possible to monitor the spreading of correlations across the
latter.

Our description can be applied to diverse systems, and we
illustrated the probing procedures for 1) a 1D long range
fermionic model, and 2) an interacting boson gas in a 2D lattice.
In the second case, in particular, we have exploited the advanced
state of the art of the cold atoms field to provide a more
detailed and experimentally feasible description. Our proposal
exemplifies the essence of the quantum probing approach, wherein
some of the typical complexity of a many-body system can be
imprinted onto the open dynamics of a smaller system, and thereby
locally extracted in a much less invasive way.

The authors would like to thank Dr. Elmar Haller and Prof. Andrew
Daley for useful discussions. This work has been supported by the
Horizon 2020 EU collaborative project QuProCS (Grant Agreement
641277).

\begin{appendix}

\section {Single Probe - Transition probability} \label {appendixA}
In this Appendix, we sketch the perturbative approach leading to
the Fermi-Golden-Rule-like equation used to describe the momentum
resolved spectroscopy protocol.  In general terms, the probability
for a ground to excited state transition, due to the interaction
with the many-body environment can be written as
\begin{equation}
\begin{split}
\Gamma_{\bar g \rightarrow \bar e}(t) = \mathrm {Tr}_{MB}\bra {\bar e} \rho (t) \ket{\bar e} =  \mathrm {Tr}_{MB}  \bra {\bar e} \hat U(t) \rho (0) \hat U^\dagger(t) \ket{\bar e},
\end{split}
\end{equation}
where $\hat U(t)$ is the time evolution operator in the
interaction picture. With the assumption of weak coupling we can
resort to an expansion in powers of $g$, so that $\hat U(t)=1
+g\hat U_1(t)+ g^2 \hat U_2(t)+...$ . Truncating up to the first
order in $g$ the relevant term is $\hat U_1(t) = - i \int_0^t \,
\mathrm dt_1 \hat H_{int} (t_1)$, so that the transition
probability takes the form
\begin{equation}
\begin{split}
\Gamma_{\bar g \rightarrow \bar e}(t) &\simeq  g^2  \mathrm {Tr}_{MB} \bra {\bar e} \hat U_1(t) \rho (0) \hat U_1^\dagger(t) \ket{\bar e} +O(g^{4}) \\
&= g^2  \int_0^t \, \mathrm dt_1    \int_0^t \, \mathrm dt_2  \\ & \mathrm {Tr}_{MB} \bra {\bar e}  \hat H_{int} (t_1) \rho (0) \hat H_{int} (t_2) \ket{\bar e}+O(g^{4}).
\end{split}
\end{equation}
In a weak coupling regime, this expression  works very well as its
first correction would be of fourth order in $g$. For the
interaction Hamiltonian in Eq.~\eqref {interaction1} we get
\begin{equation}
\begin{split}
\Gamma_{\bar g \rightarrow \bar e}(t) &\simeq  g^2  \int_0^t \, \mathrm dt_1    \int_0^t \, \mathrm dt_2   \\ & \mathrm{Tr}_{MB} \left \{ \hat \Phi {[\bar  e, \bar g]}(t_1) \rho_\beta \hat \Phi^\dagger {[\bar  g, \bar e]}(t_2)  \right \}  e^{-i \nu t_2} e^{i \nu t_1}
\\
&=  g^2  \int_0^t \, \mathrm dt_1    \int_0^t \, \mathrm dt_2 \langle  \hat \Phi^\dagger {[\bar  g, \bar e]}(t_1)  \hat \Phi {[\bar  e, \bar g]}(t_2) \rangle \\ & e^{-i \nu (t_1-t_2)}.
\end{split}
\end{equation}
In order to proceed further, we recall that $\hat \Phi{[\bar  m,
\bar n]} \propto \sum_\vk \gamma^\vk_{\bar  m, \bar n} \hat
b_\vk$, with which it is easy to obtain
\begin{equation}
\begin{split}
\Gamma_{\bar g \rightarrow \bar e}(t)
&=  g^2  \int_0^t \, \mathrm dt_1    \int_0^t \, \mathrm dt_2 \sum_{\vk, \vq} \gamma^\vk_{\bar  e, \bar g}  \gamma^{\vq *}_{\bar e, \bar g}  \langle \hat b^\dagger_\vk   b_\vq  \rangle \\& e^{i(\omega_\vk- \nu) t_1} e^{-i(\omega_\vq- \nu) t_2} \\
&= g^2  \sum_{\vk} | \gamma^\vk_{\bar  e, \bar g}|^2 n_\vk  |\int_0^t \, \mathrm dt_1  e^{i(\omega_\vk- \nu) t_1} |^2  \\ &=   g^2 t^2  \sum_{\vk} | \gamma^\vk_{\bar  e, \bar g}|^2 n_\vk   \sinc ^2[\frac {(\nu-\omega_\vk)t}{2}],
\end{split}
\end{equation}
as reported in Eq. \ref{rate2}. However, to perform full momentum
resolved spectroscopy we need to know the geometry of the system
we intend to probe. This information is crucial in order to define
the positions required for the different measurements required by
the protocol and the relations among them, which are embodied in
the functions $G_i(\vk)$.

As an example, here we calculate the functions $G_i(\vk)$ for a
simple square lattice in 2D. The $\vk$-dependent  amplitudes are
assumed to depend on an overlapping integral that involves the
probe unperturbed eigenfunctions and on the Bloch functions of the
lattice system $ \gamma^\vk_{\bar  m, \bar n}  = \int \mathrm d
\mathbf x \;  \psi_{\bar e}^* (\vx) \psi_{\bar g} (\vx) w_{\vk}
(\vx)$. The Bloch functions can, in turn, be expanded in the
Wannier basis as $w_\vk (\vx)= \sum_\vr \gamma_\vk e^{i \vk \vr}
W_\vr (\vx)$. In a 2D scenario, measurements in three different
positions are required. The  optimal basis is represented by the
following positions $ \left \{ a_0= (0,0), a_1= (1/2,0),
a_2=(1/2,1/2) \right \}$. In the first measurement  the impurity
is exactly on top of a lattice site and the locality of the
interaction allows us to consider as  relevant only the
overlapping integral that involves a single Wannier state (the one
centered on the site (0,0)). In the two other cases, the nearest
neighbours contribute equally to the overlapping integral, giving
two further contributions for the second measurement and four in
the third one. These assumptions about the localisation of the
probe allow to replace the Bloch functions in the three
overlapping integrals in the following way

\begin{equation}
\begin{split}
w_\vk (\vx)  &\simeq  \begin{cases}
 \gamma_\vk  W_{(0,0)} (\vx) \\
 \gamma_\vk (W_{(0,0)} (\vx) +e^{i \vk_x} W_{(1,0)} (\vx) ) \\
 \gamma_\vk  (W_{(0,0)} (\vx) +e^{i \vk_y} W_{(0,1)} (\vx) + \\ +e^{i \vk_x} W_{(1,0)} (\vx) +e^{i \vk} W_{(1,1)} (\vx))
\end{cases}
\end{split}
\end{equation}

For a perfect and regular lattice,  Wannier functions  centered on
different sites have the same shape. The amplitudes associated to
each transition are, then,

\begin{equation}
\begin{cases}
 |J_{0} \gamma_{\vk}!^2 \\
|J_{1} \gamma_{\vk}|^2 2[1+\cos (k_x)] \\
|J_{2} \gamma_{\vk}|^2  [16 \cos^2 (\frac {k_x a} {2} )\cos^2 (\frac {k_y a} {2} )] \\
\end{cases}
\end{equation}
%
%
where $J_{i}= \int \mathrm d \mathbf x \; \psi_{\bar e}^* (\vx-{\mathbf a_{i}}) \psi_{\bar g} (\vx-{\mathbf a_{i}}) W_{(0,0)} (\vx)$, with $i=0,1,2$.
It is important to mention that each resonant peak depends on all
the transitions with equal energy; in other words, it is a sum of
the contributions from all the possible transitions with the same
energy. This approach takes into account only the degeneracy given
by the geometry. Thanks to these considerations, we find
\begin {equation}
 \sum_{\mathbf k | \nu=\omega_{\mathbf k}} G_i(\vk)=\begin{cases}2\cos^2 (\frac {k_x a} {2} )+2\cos^2 (\frac {k_y a} {2} ) \quad i=1 \\
 16 \cos^2 (\frac {k_x a} {2} )\cos^2 (\frac {k_y a} {2}) \, \qquad i=2 \end{cases}.
\end{equation}
In 1D, just two different positions are required, and the optimal
choice is given by the probe on top of a lattice site and between
two adjacent sites. With this choice it is easy to find under the
same assumptions of a local interaction
\begin {equation}
 \sum_{ k | \nu=\omega_{\mathbf k}} G(k)=4\cos^2 (\frac {k a} {2} )
\end{equation}

\section {Two Probes - Transition probability} \label {appendixB}
In this Appendix, we derive the combined rate given in Eq.~\eqref
{rate2probes} that allows to probe spatial quantum correlations.
The starting point is to assume that the environment and the two
probes interact locally on two different sites
\begin{equation}
\hat H_{int}= \sum_{\bar n, \bar m}  \ket {\bar n}_A \bra{\bar m}_A \otimes \hat \Phi_{\vx_A}^{[\bar m_A,\bar n_A]}+\sum_{\bar n,\bar m}  \ket {\bar n}_B \bra{\bar m}_B \otimes \hat \Phi_{\vx_B}^{[\bar m_B,\bar n_B]},
\end{equation}
where $\hat  \Phi_{\vx_A}^{[m,n]}$ and  $ \hat
\Phi_{\vx_B}^{[m,n]}$ are the observables whose correlations we
want to study. We consider two entangled probes $\ket {\Psi}=\frac
{1}{\sqrt 2}(\ket {n_A ,s_B }+\ket {m_A, r_B })$, so that the
total state of the system reads
\begin{equation}
\begin{split}
\hat \rho_{tot} &=  \ket {\Psi}  \bra {\Psi} \otimes \hat \rho
\\ &=  \frac {1}{2}  (\ket {n_A, s_B }\bra {n_A ,s_B}+\ket {n_A, s_B}\bra {m_A, r_B } \\ &+\ket {m_A, r_B }\bra {n_A, s_B}+\ket {m_A ,r_B  }\bra {m_A, r_B })  \otimes \hat \rho.
\end{split}
\end{equation}
We calculate the transition probability $\Gamma ( \ket {m_A s_B})=
\bra {m_A,s_B} \hat \rho_{tot} (t) \ket {m_A,s_B}$ resorting to an
expansion of the time evolution operator to the first order in
$g$, as for the single probe case. After a few straightforward
steps, we find
\begin{equation}
\begin{split}
&\Gamma ( \ket {m_A s_B}) 
 \simeq \frac {1}{2} \int_0^t dt_1  \int_0^t dt_2  \\ & \langle  e^{i(\nu_{mA}-\nu_{nA})t_1}\hat \Phi_{\vx_A}^{[m_A,n_A]}(t_1)e^{i(\nu_{nA}-\nu_{mA})t_2}\hat \Phi_{\vx_A}^{[n_A,m_A]}(t_2)
\\&+ e^{i(\nu_{sB}-\nu_{rB})t_1} \hat \Phi_{\vx_B}^{[s_B,r_B]}(t_1)e^{i(\nu_{rB}-\nu_{sB})t_2}\hat \Phi_{x_B}^{[r_B,s_B]}(t_2)
\\&+e^{i(\nu_{mA}-\nu_{nA})t_1} \hat \Phi_{\vx_A}^{[m_A,n_A]}(t_1)e^{i(\nu_{rB}-\nu_{sB})t_2}\hat \Phi_{\vx_B}^{[r_B,s_B]}(t_2)
\\&+e^{i(\nu_{sB}-\nu_{rB})t_1} \hat \Phi_{\vx_B}^{[s_B,r_B]}(t_1)e^{i(\nu_{nA}-\nu_{mA})t_2}\hat \Phi_{\vx_A}^{[n_A,m_A]}(t_2) \rangle  \\
\end{split}
\end{equation}
Combining the result with the transition probabilities obtained in
separate single probe experiments ($\hat \rho_{tot}=  \ket {n_A}
\bra{n_A} \hat \rho$ and $\hat \rho_{tot}=   \ket {r_B} \bra{r_B}
\otimes \hat \rho$ ) we get  Eq.~\eqref{rate2probes}

\begin{equation}
\begin{split}
\overline \Gamma &=\Gamma ( \ket {m_A s_B})-\frac{1}{2}\Gamma{(\ket {m_A})}-\frac{1}{2}\Gamma{(\ket {s_B})}
\\&=\frac {g^2}{2} \int_0^t dt_1  \int_0^t dt_2  \\ & \langle   \hat \Phi_{\vx_A}^{[m_A,n_A]}(t_1)\hat \Phi_{\vx_B}^{[r_B,s_B]}(t_2) e^{i(\nu_{mA}-\nu_{nA})t_1}e^{i(\nu_{rB}-\nu_{sB})t_2}
\\ &+ \hat \Phi_{\vx_B}^{[s_B,r_B]}(t_1)\hat \Phi_{\vx_A}^{[n_A,m_A]}(t_2) e^{i(\nu_{sB}-\nu_{rB})t_1} e^{i(\nu_{nA}-\nu_{mA})t_2}\rangle.
\end{split}
\end{equation}
If the two probes are identical, and choosing
$\nu_{mA}-\nu_{nA}=\nu_{sB}-\nu_{rB}=\nu$ in order to simplify the
function $\overline \Gamma$, we have

\begin{equation}
\begin{split}
\overline \Gamma &=\frac {g^2}{2} \int_0^t dt_1  \int_0^t dt_2   \langle   \hat \Phi_{\vx_A}^{[m_A,n_A]}(t_1)\hat \Phi_{\vx_B}^{[r_B,s_B]}(t_2)
+\\ &+ \hat \Phi_{\vx_B}^{[s_B,r_B]}(t_1)\hat \Phi_{\vx_A}^{[n_A,m_A]}(t_2) \rangle e^{\nu (t_1-t_2)}.
\label{ratederivationB}
\end{split}
\end{equation}

\subsection {Density-density correlation function}
In this section, we explicitly derive the expression of $\overline
\Gamma$ for the case in which the probes and the many-body system
interact via a density-density interaction, as that introduced in
the discussion of the probing of the superfluid,
\begin{equation}
g \sum_{\bar n, \bar m} \int \mathrm d \mathbf x \;  \psi_{\bar m}^* (\vx)  \psi_{\bar n} (\vx)  \ket {\bar m} \bra {\bar n} \otimes \hat  \Phi^\dagger (\vx) \hat \Phi (\vx).
\end{equation}
Assuming a probe to be localized in the site $j$, the auxiliary
operator is defined as
\begin{equation}
 \hat \Phi_{\vx_j}^{[m, n]} =  \int \mathrm d \mathbf x \;  \psi_{\bar m}^* (\vx-\vx_j)  \psi_{\bar n} (\vx-\vx_j) \hat  \Phi^\dagger (\vx) \hat \Phi (\vx).
\end{equation}
Expanding the many-body field operator in terms of Wannier
functions,
\begin{equation}
 \hat \Phi_{\vx_j}^{[m, n]} =  \int \mathrm d \mathbf x \;  \psi_{\bar m}^* (\vx-\vx_j)  \psi_{\bar n} (\vx-\vx_j) W^*_i (\vx) W_l (\vx)  \hat  c_i^\dagger \hat
 c_l,
\end{equation}
we can take advantage of the localisation of the probe, ensuring
that the only relevant overlapping integral is that involving the
Wannier functions of the site under the probe. Therefore we get
\begin{equation}
 \hat \Phi_{\vx_j}^{[m, n]} \simeq  \gamma_{\bar n \bar m} \hat  c_j^\dagger \hat c_j  = \gamma_{\bar n \bar m} \hat n_j ,
\end{equation}
with $\gamma_{\bar n \bar m} =\int \mathrm d \mathbf x \;
\psi_{\bar m}^* (\vx-\vx_j)  \psi_{\bar n} (\vx-\vx_j) |W_j
(\vx)|^2$, and with $\hat n_j$ being the density operator for site
$j$. By inserting this result into the general form of Eq.~\eqref
{ratederivationB}, we obtain  Eq.~\eqref {density}
\begin{equation}
\overline \Gamma_{} =\frac {\bar g^2}{2} \int_0^t dt_1  \int_0^t dt_2   \langle \hat n_l(t_1)\hat n_j(t_2)+\hat n_j (t_1)\hat n_l(t_2) \rangle e^{i \nu(t_1-t_2)},
\end{equation}
where $\bar g = g \gamma_{\bar n \bar m}$.

\end{appendix}

\bibliographystyle{plain}

\end{document}